\documentclass[12pt,reqno,a4paper]{amsart}
\usepackage{amsaddr}
\usepackage{amsthm,amsmath,amssymb,eucal}
\usepackage{geometry}
\usepackage{resizegather}
\usepackage{graphicx}
\usepackage{xcolor}
\usepackage[normalem]{ulem}
\usepackage[latin1]{inputenc}
\usepackage{enumerate}
\usepackage{booktabs}

\newcommand{\ee}{\mathrm{e}}
\newcommand{\N}{\mathbb{N}}
\newcommand{\R}{\mathbb{R}}
\newcommand{\OO}{\mathcal{O}}
\newcommand{\abs}[1]{\left|{#1}\right|}

\begin{document}

\title[Vertex coupling interpolation in quantum chain graphs]{Vertex coupling interpolation in quantum chain graphs}

\author{Pavel Exner}

\address{{Doppler Institute for Mathematical Physics and Applied Mathematics, Czech Technical University, B\v rehov{\'a} 7, 11519 Prague}
{\rm and} {Department of Theoretical Physics, Nuclear Physics Institute, Czech Academy of Sciences, 25068 \v{R}e\v{z} near Prague, Czechia}
}

\email{exner@ujf.cas.cz}

\author{Jan Peka\v{r}}

\address{{Faculty of Mathematics and Physics, Charles University, V Hole\v{s}ovi\v{c}k\'ach 2, 18040 Prague}
{\rm and} {Department of Theoretical Physics, Nuclear Physics Institute, Czech Academy of Sciences, 25068 \v{R}e\v{z} near Prague, Czechia}
}

\email{honzapekar28@gmail.com}

\date{\today}

\begin{abstract}
We analyze band spectrum of the periodic quantum graph in the form of a chain of rings connected by line segments with the vertex coupling which violates the time reversal invariance, interpolating between the $\delta$ coupling and the one determined by a simple circulant matrix. We find that flat bands are generically absent and that the negative spectrum is nonempty even for interpolation with a non-attractive $\delta$ coupling; we also determine the high-energy asymptotic behavior of the bands.
\end{abstract}

\maketitle

\section{Introduction} 
\setcounter{equation}{0}

Quantum graphs, a short name for quantum mechanics of a particle the configuration space of which is a metric graph, are important both as models of various nanostructures as well as \emph{per se}, as the environment on which various quantum mechanical effects can studied using relatively simple mathematical means; we refer to the monographs \cite{BeKu13, KoNi22, Ku24} and the bibliography there.

A peculiar property of these models is that there are may ways how the wave functions can be matched at the graph vertices to produce a self-adjoint operator. In the overwhelming majority of situations, however, this multitude is reduced to graphs with wave functions continuous at the vertices leading to a one-parameter family usually referred to as \emph{$\delta$ coupling} because in case of two edges connection it is nothing else than the singular interaction formally decribed a $\delta$ function potential; the simplest among them, with the coupling parameter vanishing, is often called \emph{Kirchhoff} even if other names might be more appropriate.

Other vertex couplings, however, may be of interest too, as they may have different properties and in particular situations may fit better the physics of the problem. Motivated by the attempts to model the anomalous Hall effect \cite{StKu15, StVy23} a simple vertex coupling was proposed \cite{ExTa18} which violates the time-reversal invariance, maximally so at a fixed energy. The coupling is of a \emph{circulant type} which means that the corresponding Hamiltonian exhibits a $\mathcal{PT}$-symmetry despite being self-adjoint \cite{ExTa21}.

To understand better the consequences of time-reversal invariance violation, a family of vertex coupling interpolation between the above mentioned one and a $\delta$ coupling was constructed in \cite{ExTuTa18} and spectral properties of the corresponding quantum graph Hamiltonian were analyzed for an infinite periodic square lattice. In the present paper we are going to discuss effects of such a coupling for another graph class, namely chains of rings connected by line segments \cite{BaExTa20, BET22}. We will show that in contrast to the $\delta$ coupling, flat bands occur in the spectrum only exceptionally, and that the negative is nonempty even we interpolate with a non-attractive $\delta$ coupling. Moreover, the interpolation changes the high-energy asymptotics: the spectrum is dominated by gaps unless the chain gets degenerated by shrinking one family of the edges to points.

In the next section we describe the model in proper terms and derive the spectral condition. As a preparation we analyze in Sec.~3 the chain with the $\delta$ coupling, the main results about interpolation are presented in Sec.~4. The spectral properties we derive are numerous and we do not collect them into a single theorem; they are distinguished in the text as bullet points.

\section{The model} \label{s:model}
\setcounter{equation}{0}

\begin{figure}[b]\centering
		\includegraphics[width=.7\textwidth]{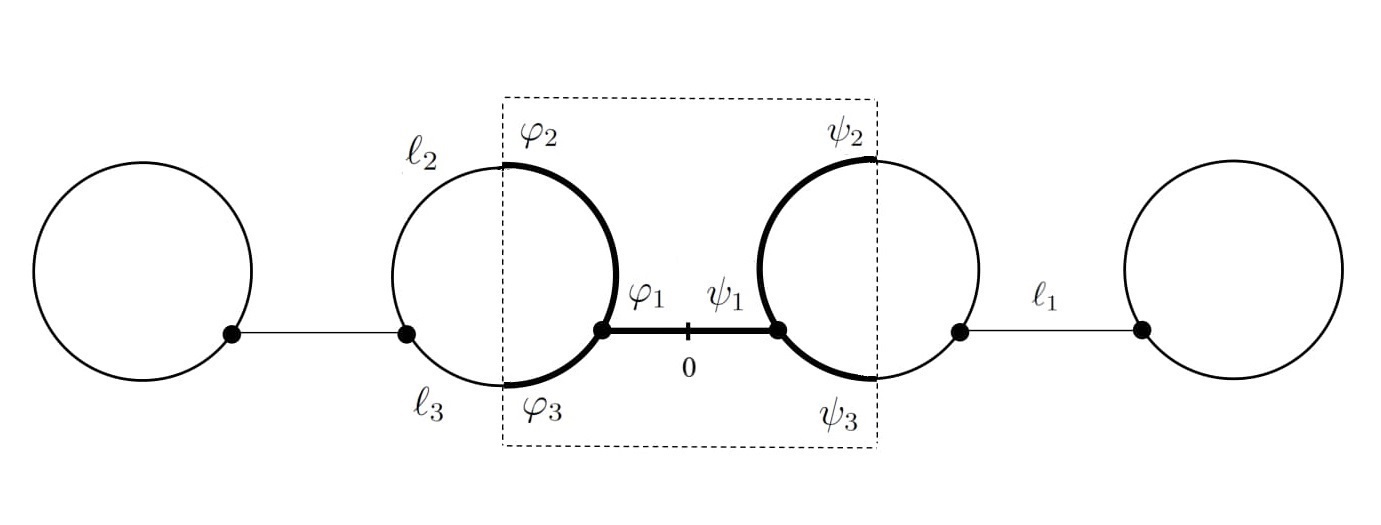}
		\caption{Periodic quantum chain; the elementary cell is highlighted \cite{BaExTa20}.}
		\label{Periodic_chain}
	\end{figure}

Working in the standard quantum graph framework \cite{BeKu13, Ku24} we consider a spinless quantum particle confined to a chain graph $\Gamma$ of the form sketched in Fig.~\ref{Periodic_chain}.
The state Hilbert space is $L^2(\Gamma)=\bigoplus_{e \in \{e\}} L^2(e)$, the direct sum over the edges of the graph, with the elements ${\Psi}=\{{\psi_l}\}$. In the absence of external forces the Hamiltonian is a multiple of the Laplacian; as the values of the physical constants are not important we put ${\hbar}=2m=1$ so that $H\{{{\psi}_l}\}=\{{-{\psi}_{l}^{''}}\}$ with the domain being a subset of $H^2(\Gamma)=\bigoplus_{e \in \{e\}} H^2(e)$. To make it a self-adjoint operator, one has to specify the condition matching the functions at the vertices. The most general choice is
	\begin{equation} \label{Boundcon}
		(U-I){\Psi}(v)+il(U+I){\Psi^{'}}(v)=0,
	\end{equation}
where $l$ is a parameter fixing the length scale and derivatives are conventionally taken in the outward direction; $U$ is a unitary $n{\times}n$ matrix with $n$ being the degree of the vertex. This conditions are usually ascribed to Kostrykin and Schrader \cite{KoSch99} and Harmer \cite{Ha00}, but they were known already to Rofe-Beketov \cite{RB69} in the context of Sturm-Liouville operators with matrix coefficients. We suppose that the chain graph is \emph{periodic} so that the coupling is the same in all the vertices.

The couplings we are interested in belong to a particular class: the matrices $U$ of \eqref{Boundcon} are \emph{circulant} \cite{Da79}, that is, of the form
    \begin{equation} \label{circulant}
		U = \begin{pmatrix} c_1 & c_2 & c_3 & {\dots} & c_n \\
			c_n & c_1 & c_2 & {\dots} & c_{n-1} \\
			{\vdots} & c_n & c_1 & {\ddots} & {\vdots} \\
			c_3 & {\dots} & {\ddots} & {\ddots} & c_2 \\
			c_2 & c_3 & {\dots} & c_n & c_1 \\
		\end{pmatrix},
	\end{equation}
being determined by the \emph{generating vector} $(c_1, c_2, c_3,{\dots}, c_n)^{\top}$. They can be diagonalized by means of the discrete Fourier transformation represented by the matrix $F$ with the entries $F_{rs}=\frac{1}{\sqrt{n}}\, \omega^{-(r-1)(s-1)}$, $r,s=1,\dots,n$, where ${\omega} := \ee^{2{\pi}i/n}$. They have all the same eigenvectors, $\phi_j =\frac{1}{\sqrt{n}} (1, {\omega}^j, {\omega}^{2j}, {\dots}, {\omega}^{(n-1)j})^T$ and their eigenvalues are $\lambda_j = \sum_{a=1}^{n} c_a{\omega}^{j(a-1)}$, $j=0,1,\dots,n-1$. Graphs with the coupling \eqref{Boundcon} of the form \eqref{circulant} exhibit a $\mathcal{PT}$-symmetry, in general a non-trivial one \cite{ExTa21}.

Two prominent examples are the \emph{$\delta$-coupling} of strength $\alpha$ with $U=U_\alpha$ corresponding to $c_j= -\delta_{j1}+\frac{2}{n+i\alpha}$, one of the most often used matching conditions, and the \emph{R-coupling} introduced in \cite{ExTa18} with $U=R$ corresponding to $c_j=\delta_{j2}$, the simplest example of a coupling violating the time-reversal invariance with interesting properties. In this paper we deal with the interpolation between the two introduced in \cite{ExTuTa18} which is given by the matching conditions \eqref{Boundcon} with the family $\{U(t): t \in [0,1]\}$ of unitary circulant matrices such that
	\begin{equation}\label{Interpol_ver_con}
		\begin{aligned}
			&U(0) = U_\alpha \; \text{and} \; U(1) = R,\\
			&t \mapsto U(t) \: \text{is continuous on} \: [0,1].\\
		\end{aligned}
	\end{equation}
Explicitly, the matrices $U(t)$ are defined through their eigenvalues chosen as
	\begin{equation}\label{Interpol_condition_eigenvalues}
		{\lambda}_j(t) = \left\{\:
		\begin{aligned}
			&\ee^{-i(1-t)\gamma} \quad &\dots\quad &j = 0, \\
			&-\ee^{i{\pi}t(\frac{2j}{n}-1)} \quad &\dots\quad &1 \le j \le n-1,\\
		\end{aligned}
		\right.
	\end{equation}
where the parameter $\gamma$ is encoding the the $\delta$-coupling strength $\alpha\in\R$,
	\begin{equation}\label{Interpol_gamma_def}
		\gamma = \arg{\frac{n+i\alpha}{n-i\alpha}} \in (-\pi,\pi) \;\; \text{so that} \;\; \frac{n-i\alpha}{n+i\alpha} = \ee^{-i\gamma}.
	\end{equation}
The choice \eqref{Interpol_condition_eigenvalues} determines the matrices $U(t)$ uniquely because the inverse discrete Fourier transform implies $c_j =\frac{1}{n}\sum_{a = 1}^{n} {\lambda}_a{\omega}^{-a(j-1)}$; for more details see \cite{ExTuTa18}.

After this preliminary, we can pass to spectral analysis of the quantum graph sketched in Fig.~\ref{Periodic_chain} with the coupling \eqref{Interpol_ver_con}. The graph is characterized by edge lengths $l_j$, $j = 1,2,3$, which we, for now, assume to be non-zero, $l_j > 0$, so that all vertices are of degree three; later we will comment on the degenerate cases where some edges shrink to points. With the obvious scaling transformation in mind, we can without loss of generality fix the loop length putting $l_2 + l_3 = 2\pi$.

The Hamiltonian of our quantum graphs, that is the Laplacian the domain of which consists of $H^2$ functions satisfying the mathching conditions \eqref{Boundcon} in the vertices with the matrices \eqref{Interpol_ver_con} will be denoted as $H_{\gamma,t}$. Since the graphs is periodic, one can employ the Floquet decomposition \cite[Chap.~4]{BeKu13},
	\begin{equation}\label{Floquet}
		H_{\gamma,t} = \int^{\oplus}_{(-t,t]} H_{\gamma,t}(\theta)\,\mathrm{d}\theta,
	\end{equation}
and reduce the analysis to inspection of the fiber operators discrete spectrum.

The elementary cell contains two vertices. In accordance with Fig.~\ref{Periodic_chain} we choose
	\begin{equation}\label{Inter_Anz}
		\begin{split}
			{\psi}_j(x)&= a_j^+\ee^{ikx} + a_j^-\ee^{-ikx},\; x \in [0, \textstyle{\frac{l_j}{2}}],\\
			{\varphi}_j(x)&= b_j^+\ee^{ikx} + b_j^-\ee^{-ikx},\; x \in [-\textstyle{\frac{l_j}{2}},0]\\
		\end{split}
	\end{equation}
with $j = 1,2,3$ as an Ansatz of the solution, where the coordinate increases from the left to the right. They have to satisfy the conditions \eqref{Boundcon} with the matrix $U(t)$ at the vertices. Furthermore, one has to impose the Floquet condition for $j = 2,3$ at the `loose' ends \cite[Sec.~4.2]{BeKu13} and couple the two parts of the connecting link at the midpoint,
	\begin{center}
		\begin{tabular}{ c c }
			${\psi}_{2}(\frac{l_2}{2}) = \ee^{i{\theta}}{\varphi}_{2}(-\frac{l_2}{2})$, & ${\psi}_{2}^{'}(\frac{l_2}{2}) = \ee^{i{\theta}}{\varphi}_{2}^{'}(-\frac{l_2}{2})$,   \\[.5em]
			${\psi}_{3}(\frac{l_3}{2}) = \ee^{i{\theta}}{\varphi}_{3}(-\frac{l_3}{2})$, & ${\psi}_{3}^{'}(\frac{l_3}{2}) = \ee^{i{\theta}}{\varphi}_{3}^{'}(-\frac{l_3}{2})$,  \\[.5em]
			${\psi}_{1}(0) = {\varphi}_1(0)$, & ${\psi}_{1}^{'}(0) = {\varphi}_{1}^{'}(0)$,
		\end{tabular}
	\end{center}
The Bloch phase factor $\ee^{i{\theta}}$ refers to the quasimomentum value $\theta \in (-\pi, \pi]$. Substituting \eqref{Inter_Anz} into these conditions we get
\begin{equation*}\label{Interpol_conditions_wf}
		\begin{aligned}
			&b_2^+ = a_2^+\ee^{ikl_2}\ee^{-i{\theta}}, \; &b_2^- = a_2^-\ee^{-ikl_2}\ee^{-i{\theta}},\\
			&b_3^+ = a_3^+\ee^{ikl_3}\ee^{-i{\theta}}, \; &b_3^- = a_3^-\ee^{-ikl_3}\ee^{-i{\theta}},\\
			&b_1^+ = a_1^+, \; &b_1^- = a_1^-.
		\end{aligned}
	\end{equation*}
To treat the $\psi$ and $\varphi$ parts of the Ansatz in a single equation system, we abuse the notation and employ the symbols $U(t)$ and $F$ also for the block-diagonal matrices {\scriptsize $\begin{pmatrix} U(t) & 0 \\ 0 & U(t)\\ \end{pmatrix}$} and {\scriptsize$\begin{pmatrix} F & 0 \\ 0 & F\\ \end{pmatrix}$}. Then the condition \eqref{Boundcon} with $U=U(t)$ becomes
	\begin{equation}\label{Interpol_con_dohromady}
		(U(t)-I) \begin{pmatrix}
			{\psi}_1(\frac{l_1}{2})\\
			{\psi}_3(0)\\
			{\psi}_2(0)\\
			{\varphi}_1(-\frac{l_1}{2})\\
			{\varphi}_2(0)\\
			{\varphi}_3(0)\\
		\end{pmatrix}
		+il(U(t)+I) \begin{pmatrix}
			-{\psi}_1^{'}(\frac{l_1}{2})\\
			{\psi}_3^{'}(0)\\
			{\psi}_2^{'}(0)\\
			{\varphi}_1^{'}(-\frac{l_1}{2})\\
			-{\varphi}_2^{'}(0)\\
			-{\varphi}_3^{'}(0)\\
		\end{pmatrix}=0,
	\end{equation}
where according to the convention the derivatives are taken in the outward direction and the numeration corresponds to the fact that both vertices have the same orientation. Relations \eqref{Inter_Anz}--\eqref{Interpol_con_dohromady} imply the equation
\begin{equation}\label{Interpol_upraveno_na_a}
		[F^*(D(t)-I)FM - klF^*(D(t)+I)FN]a=0
	\end{equation}
for the vector $a=(a_1^+,a_1^-,a_2^+,a_2^-,a_3^+,a_3^-)^T$, where
	\begin{center}
		$M =\begin{pmatrix}
			\ee^{ikl_1/2} & \ee^{-ikl_1/2} & 0 & 0 & 0 & 0 \\
			0 & 0 & 0 & 0 & 1 & 1 \\
			0 & 0 & 1 & 1 & 0 & 0 \\
			\ee^{-ikl_1/2} & \ee^{ikl_1/2} & 0 & 0 & 0 & 0 \\
			0 & 0 & \ee^{ikl_2}\ee^{-i{\theta}} & \ee^{-ikl_2}\ee^{-i{\theta}} & 0 & 0 \\
			0 & 0 & 0 & 0 & \ee^{ikl_3}\ee^{-i{\theta}} & \ee^{-ikl_3}\ee^{-i{\theta}} \\
		\end{pmatrix}$
	\end{center}
and
	\begin{center}
		$N =\begin{pmatrix}
			-\ee^{ikl_1/2} & \ee^{-ikl_1/2} & 0 & 0 & 0 & 0 \\
			0 & 0 & 0 & 0 & 1 & -1 \\
			0 & 0 & 1 & -1 & 0 & 0 \\
			\ee^{-ikl_1/2} & -\ee^{ikl_1/2} & 0 & 0 & 0 & 0 \\
			0 & 0 & -\ee^{ikl_2}\ee^{-i{\theta}} & \ee^{-ikl_2}\ee^{-i{\theta}} & 0 & 0 \\
			0 & 0 & 0 & 0 & -\ee^{ikl_3}\ee^{-i{\theta}} & \ee^{-ikl_3}\ee^{-i{\theta}} \\
		\end{pmatrix}$.
	\end{center}
The system \eqref{Interpol_upraveno_na_a} has a non-trivial solution if and only if
	\begin{equation*}
		\det[(D(t)-I)VM - kl(D(t)+I)VN] = 0,
	\end{equation*}
which is the sought spectral condition for eigenvalues $k^2$ of the fiber operator. Evaluating the determinant at the left-hand side and discarding unimportant numerical prefactors, we rewrite the condition as
	\begin{equation}\label{Interpol_fin_polynom}
		k^6\,P_6 + k^5\,P_5 + k^4\,P_4 + k^3\,P_3 + k^2\,P_2 + k\,P_1 + P_0 + k^2({\sin{\theta}}\,P_s + {\cos{\theta}}\,P_c) = 0
	\end{equation}
with the polynomial coefficients listed explicitly in Appendix~\ref{Appendix}.

Before concluding the description of the model, let us add a few remarks:

\smallskip

(i) In some situations, the condition \eqref{Interpol_fin_polynom} reduces to known results. In particular, for $t=1$ we obtain from it condition (2.8) of \cite{BaExTa20}. Furthermore, in the other extreme case, $t=0$, we get for $l_2 = l_3 = \pi$ in the limit $l_1 \to 0$ the secular equation derived in \cite{DuExTu08}, modulo the doubled value of the strength parameter $\alpha$. This is, however, easy to understand: two $\delta$-couplings converging to the same point produce a single $\delta$-coupling of the summary strength, similarly as two point interactions on the line \cite[Sec.~II.2.1]{AGHH05}

\smallskip

(ii) If $t=1$ one can use the identities $\sin{kl_3} + \sin{kl_2} = 2\sin{k{\pi}}\,\cos{k(\pi-l_3)}$ and $\cos{kl_3} - \cos{kl_2} = 2\sin{k\pi}\,\sin{k(\pi-l_3)}$ to demonstrate that there are \emph{flat bands}, or infinitely degenerate eigenvalues, coming from solutions of the spectral condition independent of quasimomentum $\theta$. The same is true for $t=0$ or $\gamma = 0$ as we will see below. Apart from those situations, both $P_s$ and $P_c$ contain terms proportional to $\sin{kl_3} + \sin{kl_2}$, but the former includes a part multiplied by $\cos{kl_3} - \cos{kl_2}$, while the latter has $\cos{kl_3} + \cos{kl_2}$ instead. This allows us to conclude that
\begin{itemize}
		\item for $t\in(0,1)$ there are no flat bands in the spectrum of $H_{\gamma,t}$, i.e. condition \eqref{Interpol_fin_polynom} has no solutions independent of $\theta$, except for $H_{0,t}$, which is considered in detail at the end of Section~\ref{s:interpolation}.
	\end{itemize}

\smallskip

(iii) We can also adopt from \cite{BaExTa20} the way to determine the spectral bands and gaps. To this aim, we rewrite condition \eqref{Interpol_fin_polynom} as
	\begin{equation}\label{Interpol_general_sol_parametrization}
		v_c\cos{\theta} + v_s\sin{\theta} = v_z
	\end{equation}
	with $v_c = k^2P_c$, $v_s = k^2P_s$, and the rest of the left-hand side constituting $v_z$. Then we introduce the angle $\vartheta$ such that $\sin{\vartheta} = \frac{v_c}{\sqrt{v^2_c + v^2_s}}$ and $\cos{\vartheta} = \frac{v_s}{\sqrt{v^2_c + v^2_s}}$; recall that in the previous point we noted that $P_c$ and $P_s$ cannot vanish simultaneously for any $t\in(0,1)$. The condition \eqref{Interpol_general_sol_parametrization} then acquires the form
	\begin{equation*}
		\sin(\theta + \vartheta) = \frac{v_z}{\sqrt{v^2_c + v^2_s}},
	\end{equation*}
and consequently, a given $k^2$ belongs to a spectral \emph{band} if
	\begin{equation}\label{Interpol_general_band_cond}
		v^2_c + v^2_s - v^2_z \ge 0,
	\end{equation}
while the corresponding spectral \emph{gap} condition reads
	\begin{equation}\label{Interpol_general_gap_cond}
		v^2_c + v^2_s - v^2_z < 0.
	\end{equation}

\section{The $\delta$-coupling} \label{s:delta}
\setcounter{equation}{0}

While the chain with the R-coupling is analysed completely, its $\delta$ counterpart was not to our knowledge discussed apart from the degenerate situation mentioned in point (i) above; it is needed to show its properties before we start discussing the interpolation. The spectral condition simplifies considerably for $t=0$, as not only $P_s=0$, but also $P_j = 0$ holds for all $j=3,4,5,6$.

\subsection*{Flat bands}

The absence of $P_s$ makes the existence of flat bands in the spectrum of $H_{\gamma,0}$ possible. Indeed, \eqref{Interpol_general_sol_parametrization} now becomes $v_c\cos{\theta} = v_z$ so a solution independent of $\theta$ exists if $v_c$ and $v_z$ vanish simultaneously. We have
	\begin{equation*}\label{v_c_param_flat}
			v_c = 64\,k^2l^2(\cos{\gamma}+1)\:\sin{k{\pi}}\:\cos{k(\pi-l_3)},
	\end{equation*}
which is zero if either $k=m$ or if $k = \frac{(2m-1){\pi}}{2(\pi-l_3)}$ holds with $m\in\mathbb{N}$; recall that in view of \eqref{Interpol_gamma_def} we have $\cos{\gamma}\ne -1$. Note also that together with the numerical prefactors in \eqref{Interpol_fin_polynom} we also have to exclude $k^2$. Instead of the Ans\"atze \eqref{Inter_Anz} we have for $k=0$ linear combinations of constant and linear functions. Should the solution be the same when multiplied by $\ee^{i\theta}$ for all $\theta$, though, it must be zero. Nevertheless, zero can belong to the spectrum as the common endpoint of two adjacent bands as we shall see below.

Choosing the first of the above possibilities, $k=m$, expression $v_z$ simplifies to
	\begin{equation*}\label{v_z_param_flat}
		\begin{aligned}
			v_z =&-16\,m^2l^2(\cos{\gamma}+1)\sin{ml_1}\,\sin{ml_2}\,\sin{ml_3}\\
			&+96\,ml\sin{\gamma}\,\cos{ml_1}\,\sin{ml_2}\,\sin{ml_3}\\
			&-144\,(\cos{\gamma}-1)\,\sin{ml_1}\,\sin{ml_2}\,\sin{ml_3},
		\end{aligned}
	\end{equation*}
which for $\sin{ml_1} \ne 0$ vanishes if
	\begin{equation*}
		\sin{ml_1}\,\sin{ml_2}\,\sin{ml_3}\,\big[ml + {\alpha}\tan(\textstyle{\frac{ml_1}{2}})\big]\big[ml - {\alpha}\cot(\textstyle{\frac{ml_1}{2}})\big] = 0\,;
	\end{equation*}
for $\sin{ml_1}=0$ the condition reduces to $\sin{\gamma}\,\sin{ml_2}\,\sin{ml_3}=0$. This implies:
	\begin{itemize}
		\item If $l_3$, and consequently $l_2$, is a rational multiple of $\pi$, $l_3 = \frac{p}{q}\pi$ with coprime $p, q \in \mathbb{N}$, then $k^2 = q^2m^2$, $m \in \mathbb{N}$, is an eigenvalue for all $p$ regardless of the other parameters.
		\item If $\gamma = 0$, or equivalently $\alpha = 0$, then the first statement is also valid for $l_j = l_1$.
		\item If $\gamma \ne 0$ and $l_1$ is an integer multiple of $\pi$, there are no flat bands except those mentioned in the first statement, if present.
		\item For other values of $l_1$, both rational and irrational, with $\gamma \ne 0$, there might be another flat band at the energy $k^2 = m^2$, as long as there exists at least one integer solution $m$ to $ml+{\alpha}\tan(\frac{ml_1}{2})=0$ or $ml-{\alpha}\cot(\frac{ml_1}{2})=0$.
	\end{itemize}

On the other hand, choosing $k = \frac{(2m-1){\pi}}{2(\pi - l_3)}$, the expression for $v_z$ can be simplified from its general form only if $\frac{(2m-1){\pi}}{2(\pi - l_3)}=m'$ with both $m,\,m'\in \mathbb{N}$; this is possible only for $l_3 = \frac{p}{q}\pi$ with coprime $p, q \in \mathbb{N}$, $q$ even. The previous conclusions then can be altered depending on the choice of $l_3$:
	\begin{itemize}
		\item If $l_3$ is a rational multiple of $\pi$ as described above, then $k^2=q^2\frac{(2m-1)^2}{4(1 - \frac{p}{q})^2}$ is an eigenvalue. Additionally, there might be another flat band $k^2=~\frac{(2m-1)^2}{4(1 - \frac{p}{q})^2}$ if there is at least one integer solution $m'$ to $m'l+{\alpha}\tan(\frac{m'l_1}{2})=0$  or $m'l-{\alpha}\cot(\frac{m'l_1}{2})=0$.
	\end{itemize}

\smallskip

The rest of the spectrum is absolutely continuous consisting of bands and gaps. In view of $v_s = 0$, the conditions \eqref{Interpol_general_band_cond} and \eqref{Interpol_general_gap_cond} simplify to
	\begin{equation}\label{Interpol_band_gap_cond_t_0}
		\abs{\frac{v_z}{v_c}} \le 1 \;\;\text{for bands}, \;\; \abs{\frac{v_z}{v_c}} > 1 \;\;\text{for gaps}.
	\end{equation}
Let us first look at the behaviour at high and low (positive) energies.

\subsection*{High energy asymptotics}

In the limit $k \to \infty$ we restrict ourselves to the leading power of $k$ in the first condition of \eqref{Interpol_band_gap_cond_t_0}, getting
	\begin{equation}\label{t_0_asymptotic}
		\frac{v_z}{v_c} = \frac{\sin{k(l_1+l_2+l_3)}-\frac{1}{2}\sin{kl_1}\sin{kl_2}\sin{kl_3}-\sin{kl_1}}{\sin{kl_2} + \sin{kl_3}} + \OO(k^{-1}).
	\end{equation}
As we have seen above, the only momentum values of $k$ where the denominator vanishes refer to flat bands, now we consider $k$ away from these points; relation \eqref{t_0_asymptotic} shows that the flat bands are generically not embedded in the continuum.

Not surprisingly, the band spectrum given by \eqref{Interpol_band_gap_cond_t_0} is asymptotically independent of the strength parameter $\gamma$.

	\begin{itemize}
		\item In the high energy limit, the spectrum of $H_{\gamma,t}$ is independent of strength parameter $\gamma$ up to an $\mathcal{O}(k^{-1})$, or $\mathcal{O}(E^{-1/2})$ error and its structure can be changed only through differences to the length parameters of a chain graph. This is easy to understand: the $\delta$ condition is an energy potential type interaction, and therefore it becomes less relevant for system behavior in the high energy regime.
	\end{itemize}

\begin{figure}[b]\centering
		\includegraphics[width=.5\textwidth]{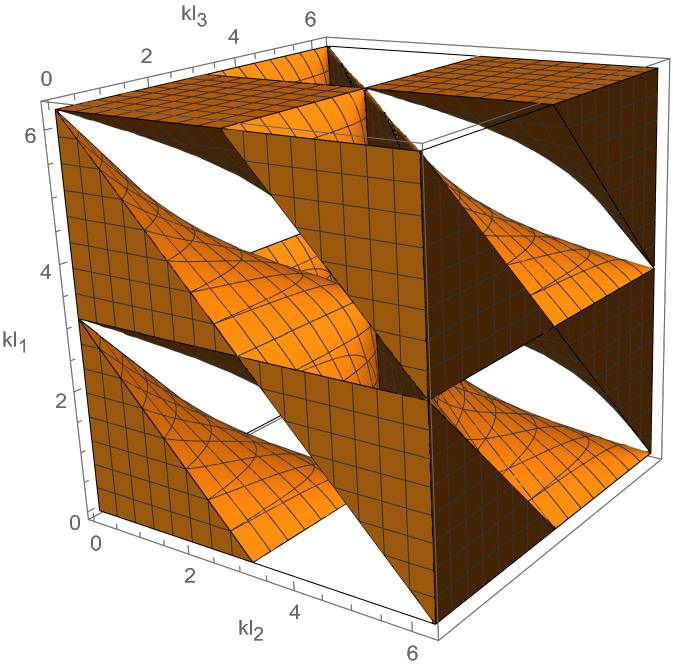}
		\caption{The volume satisfying the band condition of \eqref{Interpol_band_gap_cond_t_0}.}
		\label{Universality_plot}
	\end{figure}

While we cannot simplify \eqref{t_0_asymptotic} generally, it is possible to emphasize additional features. In particular, a quantity of interest is the probability that a randomly chosen positive number belongs to the spectrum, defined in \cite{BaBe13},
	\begin{equation}\label{Probability_in_positive}
		P_{\sigma}(H) := \lim_{E \to \infty} \frac{1}{E}\abs{\sigma(H) \cap [0,E]},
	\end{equation}
where zero can be obviously replaced by any positive number. The authors of \cite{BaBe13} noted that its value can be independent of the edge lengths as long as those are incommensurable. The reason is that the spectral condition can be understood in terms of a path on a flat torus on which points of the spectrum refer to a particular subset; if the lengths ratios are irrational, the path covers the torus in an ergodic way and the probability coincides with the relative volume of the subset. This is also the case here; as long the $l_j$'s are incommensurable, the~terms $kl_1$, $kl_2$ and $kl_3$ can be regarded as independent random variables distributed identically on interval $[0,2\pi)$, and the subset of the three-dimensional flat torus, in which the band condition \eqref{Interpol_band_gap_cond_t_0} holds, is drawn in Fig.~\ref{Universality_plot}. Note that the periodic motion displays reflection symmetry $kl_1 \leftrightarrow kl_1 - \pi$, present in \eqref{t_0_asymptotic}, taking into account \eqref{Interpol_band_gap_cond_t_0}. The only discontinuities in the surfaces of the plot are along the planes $kl_2 = 2\pi - kl_3$ and $kl_2 = kl_3 \pm \pi$ (both $\mod 2\pi$), which for our choice of scaling automatically means $k = m$ for the former and $k = \frac{(2m -1)\pi}{2(\pi-l_3)}$ for the latter, with $m \in \mathbb{N}$, i.e. instances of flat bands, which we do not consider in \eqref{t_0_asymptotic}.
The volume expression is complicated but it is not difficult to find its value numerically and dividing it by the volume $(2\pi)^3$ of the cube we get
	\begin{equation}\label{delta_probability}
		P_{\sigma}(H_{\gamma,0}) \approx 0.43.
	\end{equation}
Note also that our chain is topologically equivalent to the one in Fig.~1(e) of \cite{BaBe13}, for which the same probability value was found under the assumption that the vertex coupling is Kirchhoff; as we noted above, the difference between it and the $\delta$ coupling loses importance at high energies.

It is easy to see that the incommensurability assumption is vital. Taking, for instance, the case of maximum symmetry, $l_1 = l_2 = l_3 = \pi$, the spectral condition \eqref{t_0_asymptotic} reduces to
	\begin{equation*}
		\frac{v_z}{v_c} = -\frac{9}{4}\sin^2{k{\pi}} + 1 + \OO(k^{-1}),
	\end{equation*}
and as such the band condition is asymptotically satisfied only if
	\begin{equation*}
		0 \le \sin{k{\pi}} \le \frac{2\sqrt{2}}{3},
	\end{equation*}
and consequently, the probability can be expressed analytically as
	\begin{equation*}
		P_{\sigma}(H_{\gamma,0}) = \frac{2\arcsin{\frac{2\sqrt{2}}{3}}}{\pi} \approx 0.78
	\end{equation*}
being very different from \eqref{delta_probability}.

	\subsection*{Low energy behavior} \label{lowdelta}

	Taking Taylor expansion in the vicinity of the point $k = 0$ we get
	\begin{equation*}\label{ratio_low_energy_l_0}
		\frac{v_z}{v_c} = 1 + 3\frac{(l_2l_3 + l_1l_3 + l_1l_2)\,\tan{\frac{\gamma}{2}}}{l\,(l_2+l_3)} + \frac{9}{2}\frac{l_1l_2l_3\,\tan^2{\frac{\gamma}{2}} }{l^2\,(l_2+l_3)} + \OO(k),
	\end{equation*}
hence the band condition of \eqref{Interpol_band_gap_cond_t_0} reduces to
	\begin{equation*}
		0 \le - 3\frac{(l_2l_3 + l_1l_3 + l_1l_2)\,\tan{\frac{\gamma}{2}}}{l\,(l_2+l_3)} - \frac{9}{2}\frac{l_1l_2l_3\,\tan^2{\frac{\gamma}{2}} }{l^2\,(l_2+l_3)} \le 2.
	\end{equation*}
The question is whether the lowest positive band reaches to zero. The answer depends on length ratios; after algebraic manipulations, the idea of which is keeping $\tan{\frac{\gamma}{2}}$ as an independent variable, completing the square with respect to it, and finally taking the correct square roots given by the appropriate length ratio, we find that this happens
	\begin{equation}
		\begin{aligned}
			l_1 \le \frac{l_2l_3}{2\pi} \quad\dots\quad & \tan{\frac{\gamma}{2}} \in \big(\textstyle{-\frac{4\pi}{3}\frac{l}{l_2l_3}}, 0\big)
            \cup\big(\textstyle{-\frac{2}{3}\frac{l}{l_1}},\textstyle{-\frac{2}{3}\frac{l}{l_1} -\frac{4\pi}{3}\frac{l}{l_2l_3}}\big) \label{t_0_inequality_l1smal} \\[.3em]
            l_1 \ge \frac{l_2l_3}{2\pi} \quad\dots\quad & \tan{\frac{\gamma}{2}} \in \big(\textstyle{-\frac{2}{3}\frac{l}{l_1}}, 0\big) \cup \big(\textstyle{-\frac{4\pi}{3}\frac{l}{l_2l_3}}, \textstyle{-\frac{2}{3}\frac{l}{l_1} -\frac{4\pi}{3}\frac{l}{l_2l_3}}\big) 
		\end{aligned}
	\end{equation}
Furthermore, since $\tan(x)$ is an odd function, the spectrum threshold is positive for $\gamma > 0$ which is, of course, obvious when one inspects the quadratic form associated with the operator in question.\


	\subsection*{Negative spectrum}

The appropriate spectral condition is obtained by replacing $k$ in \eqref{Interpol_fin_polynom} by $i\kappa$, $\kappa > 0$. It is known \cite[Thm.~2.6]{BaExTa20} that if the elementary cell of a periodic graphs contains N vertices with the couplings described by unitary matrices $U_j$, $j~= 1, \dots, N$, then the negative spectrum of the corresponding Hamiltonian consists of at most $\sum_{j = 1}^{N} n_{j}^{(+)}$ bands, where $n_{j}^{(+)}$ are eigenvalues situated in the upper complex plane. In our case $N=2$ and the both matrices have all eigenvalues but one equal to $-1$, and the remaining ones are according to \eqref{Interpol_condition_eigenvalues} and \eqref{Interpol_gamma_def} in the upper half-plane if and only if $\gamma<0$.

The negative spectrum in the vicinity of $k=0$ can be treated in analogy with the positive case. In the leading term, $\sinh\,{{\kappa}l_j}$ behaves as $\sin{kl_j}$, and the same is true for $\cosh\,{{\kappa}l_j}$ and $\cos{kl_j}$. Since all the relevant parts retain their relative signs in the ratio $\frac{v_z}{v_c}$, we arrive at the condition \eqref{t_0_inequality_l1smal} again. To conclude,
	\begin{itemize}
		\item the negative spectrum is non-empty if and only if the coupling is attractive, $\gamma<0$, in which case it consists of at most two negative bands,
		\item the first positive and negative band are either separated from $k=0$ or extend to this point simultaneously.
	\end{itemize}

	\subsection*{Degenerate cases: vertices of degree four}

Finally, let us look at the situation when one group of edges length tends to zero giving rise to a graph with vertices of degree four, assuming either $l_1 \to 0$, or, without the loss of generality, $l_3 \to 0$ (not the limits together, of course). In~both cases, the spectral condition does not simplify much. Furthermore, the claims about flat bands made above in the general case remain valid whenever relevant; it is not surprising since the corresponding eigenfunctions can be chose compactly supported with zero values at the vertices.

Consider first the limit $l_1 \to 0$. By a straightforward computation we find that in the leading order the  spectral condition \eqref{Interpol_general_band_cond} reads
	\begin{equation*}
		(\sin{kl_2} + \sin{kl_3})^2 - \sin^2{k(l_2+l_3)}
			= 4\sin{kl_2}\,\sin{kl_3}\,\sin^2{k\pi} \ge 0
	\end{equation*}
 up to an $\OO(k^{-1})$ error, which yields the condition,
	\begin{equation}\label{l_1_0_asymptotic}
		\sin{kl_2}\,\sin{kl_3} \ge 0,
	\end{equation}
the same as obtained in \cite{BaExTa20} for the limit $l_1 \to 0$ in case of the other extreme coupling, $t=1$. We can also get it from \eqref{t_0_asymptotic} simply by setting $l_1 = 0$ and returning to the form \eqref{Interpol_general_band_cond} of the spectral band condition, i.e. by taking the square of the given fraction with additional algebraic manipulations. The probability \eqref{Probability_in_positive} then depends on the values of $l_2$ and $l_3$:
	\begin{itemize}
		\item For the tightly connected mirror-symmetric chain, $l_1=0$ and $l_2=l_3=\pi$, we have $P_{\sigma}(H_{\gamma,t}) = 1$, while $P_{\sigma}(H_{\gamma,t}) = \frac12$ holds whenever $l_2$ and $l_3$ are unequal (and non-zero).
	\end{itemize}
The first claim is obvious because the right-hand side of \eqref{l_1_0_asymptotic} is then non-negative. What is more interesting is the asymmetric situation, recalling what we said above about the Band-Berkolaiko universality \cite{BaBe13}. In this case again the points of the spectrum correspond to the values of $kl_2$ and $kl_3$ belonging to a particular subset of the torus $(0,2\pi)^2$. If the edges are incommensurate, the trajectory corresponding to $k$ running to infinity fills the torus uniformly as a consequence of the ergodic theorem, and the probability is thus given by the relative area of the subset. Here, however, the subset has a particularly simple shape, consisting of two squares, $(0,\pi)^2\cup(\pi,2\pi)^2$ with the area one half of that of the torus, and the result holds even if $l_2$ and $l_3$ are rationally related, as long as they are unequal.

Let us pass to the lower part of the tight array spectrum. The threshold is obviously positive for $\gamma>0$, and furthermore
	\begin{itemize}
		\item the lowest positive band extends by \eqref{t_0_inequality_l1smal} to $k = 0$ provided
		\begin{equation*}\label{l_1_0_connected_to_zero}
			0 \ge \tan{\frac{\gamma}{2}} \ge -\frac{4\pi}{3}\frac{l}{l_2l_3}\,;
		\end{equation*}
the same is true for the negative spectral band which is now single only due to the general result mentioned above because the elementary cell contains in this case just one vertex.
	\end{itemize}

Let us turn to the other degenerate case, $l_3 \to 0$. The quantity $v_c$ becomes
	\begin{equation*}
		\begin{aligned}
			v_c &= 32\,k^2l^2(\cos{\gamma}+1)\,\sin{2k{\pi}}.
		\end{aligned}
	\end{equation*}
As before, all the values $k^2=m^2$ with $m \in \mathbb{N}$ belong to the positive spectrum, but additionally we can have flat bands also at the squares of half-integer numbers, $k=\frac{2m-1}{2}$. Indeed, in the absence of $v_s$ the condition \eqref{Interpol_general_sol_parametrization} for these $k$ requires
	\begin{equation*}
		v_z=64\bigg{(}\frac{2m-1}{2}\bigg{)}^2l^2(\cos{\gamma}+1)\sin{\frac{(2m-1)\pi}{2}}\,\cos{\frac{(2m-1)(\pi+l_1)}{2}} = 0,
	\end{equation*}
 and therefore
	\begin{itemize}
		\item flat bands are for $l_3 \to 0$ present if $l_1$ is a $\pi$-multiple of an even integer (for all $m\in\mathbb{N}$), or more generally, if $l_1=\frac{p}{q}\pi$ with coprime $p$ and $q$, and $(2m-1)(1+\frac{p}{q}) = 1\;(\!\!\!\!\mod 2)$ (for some $m\in\mathbb{N}$).\\
	\end{itemize}
In the high energy limit, we have from \eqref{Interpol_general_band_cond}, or from \eqref{t_0_asymptotic} similarly to the previous case,
	\begin{equation*}
		4\sin^2{\frac{kl_2}{2}}\,\sin{kl_1}\,\sin{k(l_2+l_1)}\ge 0,
	\end{equation*}
and since the first factor is non-negative, it is equivalent to
	\begin{equation*}
		\sin{kl_1}\,\sin{k(2\pi+l_1)} \ge 0,
	\end{equation*}
up to $\mathcal{O}(k^{-1})$, which is once again the same condition as was found by \cite{BaExTa20} for $t=1$ in the limit $l_3\to 0$. Even though there is a periodicity if $l_1$ is a rational multiple of $\pi$, using argument similar to when $l_1$ was approaching 0, it is true that
	\begin{itemize}
		\item for any value of length $l_1$ in the limit $l_3 \to 0$
		\begin{equation*}
			P_{\sigma}(H_{\gamma,t}) = \frac{1}{2}.
		\end{equation*}
	\end{itemize}
	When $l_1 \to \infty$, the positive spectrum becomes 'denser' if we talk about the number of spectral bands (or gaps) per fixed interval, but the percentage of the momentum scale covered remains the same.
	\begin{itemize}
		\item Low energy behaviour for $l_3 \to 0$ can be again simply described by one condition
		\begin{equation*}\label{l_3_0_connected_to_zero}
			0 \ge \tan{\frac{\gamma}{2}} \ge -\frac{2}{3}\frac{l}{l_1}.
		\end{equation*}
		Whether is the positive and negative spectrum connected to zero is determined (up to the exclusion of 0 from the latter) by this condition, and the negative spectrum still has at most one spectral band.
	\end{itemize}

\section{Interpolation} \label{s:interpolation}
\setcounter{equation}{0}

\begin{figure}[b]\centering
		\includegraphics[width=.5\textwidth]{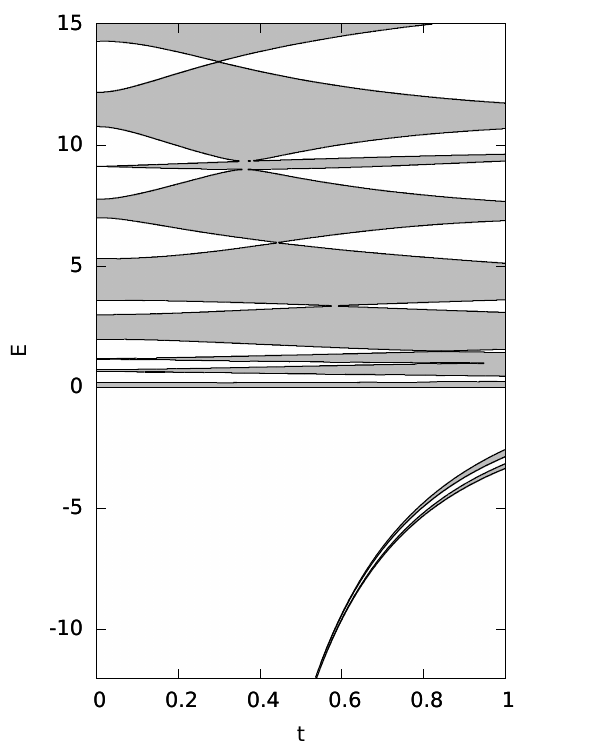}
		\caption{The band-gap structure dependence on interpolation parameter $t$ for a quantum chain with $l = 1$, $l_1 = 2$, $l_3 = \frac{\pi}{2}$ and $\gamma = 0$.}
		\label{Interpol_plot_Kirch}
	\end{figure}

\begin{figure}[b]\centering
		\includegraphics[width=.5\textwidth]{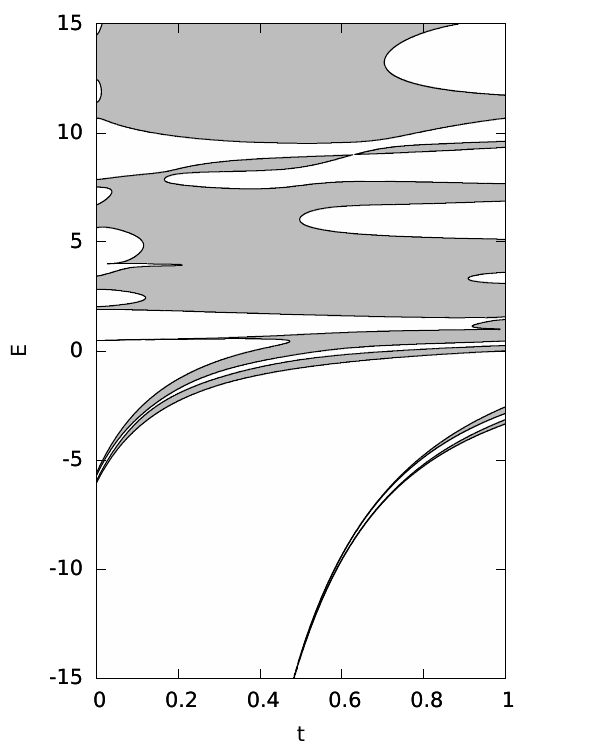}
		\caption{The same as Fig. \ref{Interpol_plot_Kirch}, but with $\gamma = -\frac{3\pi}{4}$.}
		\label{Interpol_plot_neg_gam}
	\end{figure}

After this long preliminary, let us focus on our proper topic, the interpolation between the two types of the vertex coupling given by \eqref{Interpol_ver_con} and \eqref{Interpol_condition_eigenvalues}, certainly nontrivial as the two couplings give rise to a substantial different spectral behavior. At the same time, the extreme cases, $t = 0$ and $t = 1$, are the only ones is which the general spectral condition \eqref{Interpol_fin_polynom} simplifies considerably due to vanishing of some of the polynomial coefficients. We have already shown that for $t\in(0,1)$ there are \emph{no flat bands}, hence we have to consider the absolutely continuous spectral bands only. Their dependence on the interpolation parameter is illustrated in Figs.~\ref{Interpol_plot_Kirch} and~\ref{Interpol_plot_neg_gam}, with $t=0$ corresponding to the Kirchhoff and an attractive $\delta$ coupling, respectively. Let us inspect some shared aspects of the spectral bands behavior.

	\subsection*{High-energy asymptotics}

In the leading order of the momentum variable $k$, the gap condition \eqref{Interpol_general_gap_cond} becomes
 \begin{equation*}
		-1296\:k^{12}\:l^{12}\:(\cos{\gamma(1-t)}+1)^2\:{\sin}^2{kl_1}\:{\sin}^2{kl_2}\:{\sin}^2{kl_3} + \mathcal{O}(k^{11}) < 0,
\end{equation*}
taken into account that the positive factor $(\cos{\frac{{\pi}t}{3}}-1)^2$ may be neglected. Consequently, the spectrum is at high energies dominated by gaps; spectral bands may occur only in the vicinity of the points
 \begin{equation}\label{Interpol_body_t_ne_0}
		k_{m,j} = \frac{m{\pi}}{l_j}, \quad m\in\mathbb{N}, \; j = 1,2,3.
 \end{equation}
To find the asymptotic behavior of the bands, we rewrite \eqref{Interpol_fin_polynom} as
	\begin{equation}\label{Interpol_asymptotic_t_ne_0}
		\frac{P_6}{l^6} + \frac{P_5}{kl^6} + \frac{P_4}{k^2l^6} + \frac{{P_c\:{\cos}\:{\theta}+P_s\:{\sin}\:{\theta}}}{k^4l^6} = \mathcal{O}(k^{-3}).
	\end{equation}
The analysis in each of the three cases of \eqref{Interpol_body_t_ne_0} is similar; in view of the symmetry with respect the swap of $l_2$ and $l_3$ we will write it explicitly only for $l_1$ and $l_3$.

Consider first $l_j = l_1$ and expand the polynomial coefficients in the vicinity of the points \eqref{Interpol_body_t_ne_0}, $k=\frac{\pi m}{l_1}+\delta$; plugging them into \eqref{Interpol_fin_polynom} we get the condition
 \begin{equation}\label{Asymptotic_expression_t_ne_0_k_inf}
		\begin{aligned}
			\Big\{&-(-1)^m\frac{24{\beta}_b{l_1}}{m{\pi}l}\sin{\textstyle{\frac{m{\pi}l_2}{l_1}}}\sin{\textstyle{\frac{m{\pi}l_3}{l_1}}}\\		 &+\frac{(-1)^m{l^2_1}}{m^2{\pi}^2l^2}\big[\cos{\textstyle{\frac{m{\pi}l_2}{l_1}}}\sin{\textstyle{\frac{m{\pi}l_3}{l_1}}}
+\sin{\textstyle{\frac{m{\pi}l_2}{l_1}}}\cos{\textstyle{\frac{m{\pi}l_3}{l_1}}}\big][9{\beta}_d + {\beta}_e]\\
			 &+\frac{16{l^2_1}}{m^2{\pi}^2l^2}\big[\sin{\textstyle{\frac{m{\pi}l_2}{l_1}}}+\sin{\textstyle{\frac{m{\pi}l_3}{l_1}}}\big][-{\beta}_c{\cos}{\theta} + {\beta}_f{\sin}{\theta}]\Big\}\\
			+{\delta}\Big\{&(-1)^m{l_1}\sin{\textstyle{\frac{m{\pi}l_2}{l_1}}}\sin{\textstyle{\frac{m{\pi}l_3}{l_1}}}\big[36{\beta_a} +\frac{3{l^2_1}}{m^2{\pi}^2l^2}(-3{\beta_d}+\beta_e)\big]\\
			&-(-1)^m\frac{24{\beta}_b{l_1}}{m{\pi}l}\big[(l_1+l_3)\sin{\textstyle{\frac{m{\pi}l_2}{l_1}}}\cos{\textstyle{\frac{m{\pi}l_3}{l_1}}} + (l_1+l_2)\cos{\textstyle{\frac{m{\pi}l_2}{l_1}}}\sin{\textstyle{\frac{m{\pi}l_3}{l_1}}}\big]\\
			&+\frac{(-1)^m{l^2_1}}{m^2{\pi}^2l^2}\big[(l_1+l_2+l_3)\cos{\textstyle{\frac{m{\pi}l_2}{l_1}}}\cos{\textstyle{\frac{m{\pi}l_3}{l_1}}} - (l_2 + l_3)\sin{\textstyle{\frac{m{\pi}l_2}{l_1}}}\sin{\textstyle{\frac{m{\pi}l_3}{l_1}}}\big] (9{\beta_d} + {\beta_e})\\
			&-\frac{(-1)^m16{l^2_1}}{m^2{\pi}^2l^2}l_1{\beta_c}
			+\frac{16{l^2_1}}{m^2{\pi}^2l^2}\big[l_2\cos{\textstyle{\frac{m{\pi}l_2}{l_1}}} + l_3\cos{\textstyle{\frac{m{\pi}l_3}{l_1}}}\big][-{\beta}_c{\cos}{\theta} + {\beta}_f{\sin}{\theta}]\Big\}\\			
+{\delta}^2\Big\{
            &(-1)^m36{\beta_a}\big(l_1l_3\sin{\textstyle{\frac{m{\pi}l_2}{l_1}}}\cos{\textstyle{\frac{m{\pi}l_3}{l_1}}}+l_1l_2\sin{\textstyle{\frac{m{\pi}l_3}{l_1}}}
\cos{\textstyle{\frac{m{\pi}l_2}{l_1}}}\big)\\
			 &-(-1)^m\frac{24{\beta}_b{l_1}}{m{\pi}l}\big[(l_1l_2+l_2l_3+l_1l_3)\cos{\textstyle{\frac{m{\pi}l_2}{l_1}}}\cos{\textstyle{\frac{m{\pi}l_3}{l_1}}}\\			 &-\big(\textstyle{\frac12}l^2_1+l_1l_2+\textstyle{\frac12}l^2_2+l_1l_3+\textstyle{\frac12}l^2_3\big)\sin{\textstyle{\frac{m{\pi}l_2}{l_1}}}\sin{\textstyle{\frac{m{\pi}l_3}{l_1}}}\big] + \mathcal{O}(m^{-2})\Big\}\\
			+{\delta}^3&\Big\{(-1)^m36{\beta_a}\big[l_1l_2l_3\cos{\textstyle{\frac{m{\pi}l_2}{l_1}}}\cos{\textstyle{\frac{m{\pi}l_3}{l_1}}}\\			 &-\big(\textstyle{\frac16}l^3_1+\textstyle{\frac12}l_1l^2_2+\textstyle{\frac12}l_1l^2_3\big)\sin{\textstyle{\frac{m{\pi}l_2}{l_1}}}\sin{\textstyle{\frac{m{\pi}l_3}{l_1}}}\big]+\mathcal{O}(m^{-1})\Big\} +\mathcal{O}({\delta}^4) = 0,
		\end{aligned}
	\end{equation}
regarded as an equation determining $\delta$ as a function of $\theta$, where
	\begin{equation*}\label{Interpolating_betas}
		\begin{aligned}
			{\beta}_a &:= \cos{\gamma(1-t)} + 1,\\
			{\beta}_b &:= \sin{\gamma(1-t)},\\
			{\beta}_c &:= \cos{\gamma(1-t)}\;\frac{\cos{\frac{{\pi}t}{3}}+2}{\cos{\frac{{\pi}t}{3}}-1} + 2\:\frac{\cos{\frac{{\pi}t}{3}}+\frac{1}{2}}{\cos{\frac{{\pi}t}{3}}-1},\\
			{\beta}_d &:= \cos{\gamma(1-t)}\;\frac{\cos{\frac{{\pi}t}{3}}+3}{\cos{\frac{{\pi}t}{3}}-1} + 3\:\frac{\cos{\frac{{\pi}t}{3}}+\frac{1}{3}}{\cos{\frac{{\pi}t}{3}}-1},\\
			{\beta}_e &:= 7\:\cos{\gamma(1-t)}\;\frac{\cos{\frac{{\pi}t}{3}}+\frac{5}{7}}{\cos{\frac{{\pi}t}{3}}-1} + 5\:\frac{\cos{\frac{{\pi}t}{3}}+\frac{7}{5}}{\cos{\frac{{\pi}t}{3}}-1},\\
			{\beta}_f &:= \sqrt{3}\:\sin{\gamma(1-t)}\;\frac{\sin{\frac{{\pi}t}{3}}\:(\cos{\frac{{\pi}t}{3}+1})}{(\cos{\frac{{\pi}t}{3}}-1)^2}.\\
		\end{aligned}
	\end{equation*}
In the high-energy asymptotic regime, the dispersion function branch $E_{m}(\theta)$ around $k=\frac{m{\pi}}{l_j}$, in dependence on the Bloch factor, is
	\begin{equation*}\label{E_high_E_limit}
		E_{m}(\theta) := k^2_{m,j}(\theta) = \big{(}\frac{m\pi}{l_j}\big{)}^2 + 2\,\frac{m\pi}{l_j}\delta(\theta) + \mathcal{O}(m^{0})
	\end{equation*}
as $m\to\infty$. To assess the \emph{width of the $m$-th spectral band} we write it as
 \begin{equation}\label{Delta_E_high_E_limit}
			{\Delta}E_m =  {\lvert}{k^2_{m,j}}({\theta}_{max}) - {k^2_{m,j}}({\theta}_{min}){\rvert}
			\approx 2\,\frac{m\pi}{l_j}{\lvert}\delta({\theta}_{max})) -\delta({\theta}_{min}){\rvert},
\end{equation}
where ${\theta}_{max}$ is the point of the Brillouin zone corresponding to the maximun of the dispersion curve, and analogously for ${\theta}_{min}$; the symbol $\approx$ means the value up to an error term of the next order in $m$. Note that since the vertex coupling given by \eqref{Interpol_condition_eigenvalues} violates the time-reversal invariance unless $t=0$, the dispersion functions need be neither symmetric nor anisymmetric with respect to the center of the Brillouin zone \cite{EKW10}.

Assuming that both $l_2$ and $l_3$ are incommensurate with $l_1$, one can obtain $\delta(\theta)$ in the asymptotic regime by retaining only the leading term in $m$ in the linear part of the $\delta$ expansion. Keeping then only the terms that do not vanish in the subtraction of \eqref{Delta_E_high_E_limit}, we get
	\begin{equation*}
		\delta(\theta) \approx \frac{4(-1)^m{l_1}}{9m^2{\pi}^2l^2} \frac{[\sin{\frac{m{\pi}l_2}{l_1}}+\sin{\frac{m{\pi}l_3}{l_1}}][{\beta}_c{\cos}{\theta} - {\beta}_f{\sin}{\theta}]}{{\beta_a}\sin{\frac{m{\pi}l_2}{l_1}}\sin{\frac{m{\pi}l_3}{l_1}}}\,;
	\end{equation*}
note that $\beta_a$ never vanishes. Finding the maxima and minima of the dispersion curves is equivalent to finding the extrema of $\delta(\theta)$. Differentiating with respect to $\theta$, we find $-\frac{\beta_f}{\beta_c} = \tan{\theta}$ which has two solutions on $(-{\pi},\pi]$ the difference of which is $\pi$. Observing further that shifting the argument of sine and cosine by $\pi$ means switching the sign, we get from \eqref{Delta_E_high_E_limit} the expression
	\begin{equation}\label{Band_width_irrational}
		\begin{aligned}
			{\Delta}E_m =\frac{16}{9m{\pi}l^2}{\bigg|}\frac{\sqrt{{\beta_c}^2 + {\beta_f}^2}(\sin{\frac{m{\pi}l_2}{l_1}}+\sin{\frac{m{\pi}l_3}{l_1}})}{{\beta_a}\sin{\frac{m{\pi}l_2}{l_1}}\sin{\frac{m{\pi}l_3}{l_1}}}{\bigg|} + \mathcal{O}(m^{-2}),
		\end{aligned}
	\end{equation}
where we used the fact that
	\begin{equation*}
		\begin{aligned}
			-{\beta}_c\cos(\arctan({-\frac{\beta_f}{\beta_c}}))+{\beta}_f\sin({\arctan({-\frac{\beta_f}{\beta_c}}})) = -\sqrt{{\beta_c}^2 + {\beta_f}^2}.
		\end{aligned}
	\end{equation*}

If at least one of the fractions $\frac{l_3}{l_1}$ or $\frac{l_2}{l_1}$ is rational, there exist values of $m$ for which $\sin{\frac{m{\pi}l_2}{l_1}}$ or $\sin{\frac{m{\pi}l_3}{l_1}}$, and consequently the denominator of \eqref{Band_width_irrational}, vanish. This forces us to go to ${\delta}^2$ terms in the expansion \eqref{Asymptotic_expression_t_ne_0_k_inf}. If $l_2 \ne l_3$, let us assume that for a given $m$ it holds
 \begin{equation} \label{ratcond}
		\sin{\frac{m{\pi}l_2}{l_1}} = 0, \quad \cos{\frac{m{\pi}l_2}{l_1}} = (-1)^{ml_2/l_1},
	\end{equation}
while $\sin{\frac{m{\pi}l_3}{l_1}}$ does not vanish. Two solutions of the resulting quadratic equation again yield the same band width, so it is sufficient to write the leading term of one of them,
 \begin{equation}\label{Irreg_discriminant}\small
		\delta(\theta) \approx \sqrt{\bigg{(}\frac{{\beta}_b{l_1}}{3m{\pi}{\beta_a}l}\bigg{)}^2\frac{(l_1+l_2)^2}{l^2_1l^2_2}-\frac{l^2_1[9{\beta}_d + {\beta}_e+(-1)^{m(1+l_2/l_1)}16(-{\beta}_c{\cos}{\theta} + {\beta}_f{\sin}{\theta})]}{36{m^2{\pi}^2l^2{\beta_a}l_1l_2}}}.
	\end{equation}
Looking for extrema in the Brillouin zone, we have to make sure that the argument of the square root in \eqref{Irreg_discriminant} is non-negative. This is obviously true for its first term and for the denominator of the second one. Furthermore, the function $\theta\mapsto (-1)^{m(1+l_2/l_1)}(-{\beta}_c{\cos}{\theta} + {\beta}_f{\sin}{\theta})$ is sign changing and its minimum is certainly negative. The term in question is $9{\beta}_d + {\beta}_e$ which is easily seen to be always negative. Seeking its extrema, we have
 \begin{equation*}
		\frac{\partial}{{\partial}{\gamma}}(9{\beta}_d + {\beta}_e) = \frac{(1-t)\sin{\gamma(1-t)}}{1-\cos{\frac{{\pi}t}{3}}}16\big(2+\cos{\textstyle{\frac{{\pi}t}{3}}}\big) = 0,
	\end{equation*}
requiring, on intervals we consider, either $t=1$ or $\gamma=0$. In the former case we have $9{\beta}_d + {\beta}_e = -144$ regardless of the chosen $\gamma$. The other partial derivative, on the line $\gamma=0$, reads
	\begin{equation*}
		\frac{\partial}{{\partial}{t}}(9{\beta}_d + {\beta}_e) = 32\frac{{\pi}\sin{\frac{{\pi}t}{3}}}{(1-\cos{\frac{{\pi}t}{3}})^2} = 0,
	\end{equation*}
which for $t\ne0$ does not have a solution. Thus no local extrema exist for $t\in(0,1)$, and a check of the interval endpoints shows that $-144$ is in fact the maximum value. This means that \eqref{Irreg_discriminant} has always at least one real solution. If there are two, the band width is determined by their difference. If there is just one, band width is given by it; note that the argument of square root is an analytic function of $\theta$, so there must be a $\theta$ for which it is vanishes, marking the other edge of this spectral band. In both cases, however, \eqref{Irreg_discriminant} implies $\delta(\theta) \propto m^{-1}$, hence it follows from \eqref{Delta_E_high_E_limit} that
  \begin{equation*}
		{\Delta}E_m = \text{const}+\mathcal{O}(m^{-1})\quad\text{as}\,\; m\to\infty,
	\end{equation*}
where the constant depends on the edge lengths and the coefficients $\beta_j$ involved, and a similar conclusion if the roles of $l_2$ and $l_3$ are swaped; note that if \eqref{ratcond} is satisfied for some $m$, it hold also for $m+\frac{l_1}{l_2}n$ for any $n\in\N$.

Finally, it may happen that for a certain $m$, and therefore for an infinite sequence of indices, $\sin{\frac{m{\pi}l_2}{l_1}}$ and $\sin{\frac{m{\pi}l_3}{l_1}}$ vanish simultaneously, for instance, if all the $l_j$'s are rational multiples of $\pi$. In that case we have to use the expansion \eqref{Asymptotic_expression_t_ne_0_k_inf} to the third order in $\delta$ noting that $-16\beta_c+9\beta_d+\beta_e=0$; we get
	\begin{equation}\label{Determinant_posledni}
		\begin{aligned}
			\delta(\theta) \approx	 \bigg{\{}\bigg{(}&\frac{{\beta}_b{l_1}}{3m{\pi}{\beta_a}l}\bigg{)}^2\frac{(l_1l_2+l_2l_3+l_1l_3)^2}{l^2_1l^2_2l^2_3}\\
			&-\frac{l^2_1(l_2+l_3)[9{\beta}_d + {\beta}_e+(-1)^{m(1+\frac{l_2}{l_1})}16(-{\beta}_c{\cos}{\theta} + {\beta}_f{\sin}{\theta})]}{36{m^2{\pi}^2l^2{\beta_a}l_1l_2l_3}}\bigg{\}}^{1/2}.
		\end{aligned}
	\end{equation}
By the same arguments as above, we arrive at
	\begin{equation*}
		{\Delta}E_m = \text{const}+\mathcal{O}(m^{-1})
	\end{equation*}
with the leading term dependent on the parameters involved. Summarizing the discussion, the spectral bands of $H_{\gamma,t}$ with $t\in(0,1)$ located in the vicinity of the points $\frac{m\pi}{l_1}$ have the following properties:
	\begin{itemize}
		\item if both $l_2$ and $l_3$ are incommensurate with $l_1$, their width behaves asymptotically as $\mathcal{O}(k^{-1})$ for $k \to \infty$,
		\item if $l_2$ or $l_3$ is commensurate with $l_1$, there exists an infinite sequence of bands of an asymptotically constant width, the value of which depends on $t$, $\gamma$, $l$ and $l_j,\,j=1,2,3$; the rest of these bands follows the  $\mathcal{O}(k^{-1})$ asymptotics.
	\end{itemize}

The spectral bands located in the vicinity of the points $\frac{m\pi}{l_3}$ can be treated in an analogous way. If both the $l_1$ and $l_2$ are incommensurate with $l_3$, one uses expansion \eqref{Interpol_asymptotic_t_ne_0} up to the term linear in $\delta$ and arrives at
	\begin{equation*}\label{Band_width_irrational_l_3}
		\begin{aligned}
			{\Delta}E_m = \mathcal{O}(m^{-1}),
		\end{aligned}
	\end{equation*}
If $\frac{l_1}{l_3}$ is rational and $m$ such that  $\sin{\frac{m{\pi}l_1}{l_3}} = 0$ with $\sin{\frac{m{\pi}l_2}{l_3}}$ nonzero, we have to go up to $\delta^2$ term. The obtained $\delta(\theta)$ is similar to \eqref{Irreg_discriminant} leading to
	\begin{equation*}
		\begin{aligned}
			{\Delta}E_m = \text{const}+\mathcal{O}(m^{-1})
		\end{aligned}
	\end{equation*}
in the leading order. On the other hand, if $\frac{l_2}{l_3}$ is rational, $\sin{\frac{m{\pi}l_2}{l_3}} = 0$, and  $\sin{\frac{m{\pi}l_1}{l_3}}$ is nonzero, the obtained equation and its solutions for $\delta(\theta)$ depends on the parity of $m$, but irrespective of it we find
	\begin{equation*}
		{\Delta}E_m = \mathcal{O}(m^{-1})
	\end{equation*}
Finally, if $\sin{\frac{m{\pi}l_1}{l_3}}$ and $\sin{\frac{m{\pi}l_2}{l_3}}$ vanish simultaneously for a sequence of the indices $m$, the solution we get for $\delta(\theta)$ by taking the expansion to the $\delta^3$ term is similar to \eqref{Determinant_posledni} and
	\begin{equation*}
			{\Delta}E_m = \text{const}+\mathcal{O}(m^{-1})
	\end{equation*}
with the constant dependent on the parameters involved. Considering finally the situation in which $l_2$ and $l_3$ are swapped, we can conclude that
	\begin{itemize}
		\item the asymptotics of the bands located near the points $k=\frac{m\pi}{l_3}$ is similar to that discussed above: their widths behaving as $\mathcal{O}(k^{-1})$ whenever $l_1$ and $l_2$ are incommensurate with $l_3$, while if $l_1$ and $l_3$ are commensurate there are band sequences of asymptotically constant width,
		\item the analogous claims applies to the bands located in the vicinity of $k = \frac{m\pi}{l_2}$.
	\end{itemize}
The positive energy spectrum of $H_{\gamma,t}$ thus consists of three sequences of bands around the points $k^2 = \big(\frac{m\pi}{l_j}\big)^2$ which may in general overlap. They either shrink as $k\to\infty$ or have asymptotically constant widths depending on $t$, $\gamma$, $l$ and $l_j,\,j=1,2,3$. Consequently, their total measure in an interval $[0,k^2]$ is bound by a multiple of $k$ which, in particular, implies for the probability \eqref{Probability_in_positive} the following result:
	\begin{itemize}
		\item chain graph with vertices of degree three has $P_{\sigma}(H_{\gamma,t}) = 0$ for any $t\in(0,1]$.
	\end{itemize}
We have included here the case $t=1$ in which the result can be found in \cite[eq.~(2.27)]{BET22}. On the other hand, we know from \eqref{delta_probability} that $P_{\sigma}(H)$ is nonzero for $t=0$. Combining Floquet decomposition \eqref{Floquet} with the result of \cite{BLS19} we find that the spectrum of $H_{\gamma,t}$ converges in the sense of Hausdorff set distance as $t\to 0+$; the comparison of the probabilities shows, however, that the convergence can be uniform on bounded intervals only.

\medskip

One may ask whether the point $k=0$ belongs to the spectrum in analogy with the situation discussed in the appropriate part of Sec.~\ref{s:delta}, however, the condition coming from Taylor expansion of the coefficients in \eqref{Interpol_fin_polynom} is complicated and does not allow for any reasonable insight, hence we do not present it.

	\subsection*{Negative spectrum}

Concerning the negative spectrum of $H_{\gamma,t}$, we are once again interested in two particular characteristics, namely the number of spectral bands, and whether the lowest one, if~present, is connected to $k^2=0$. The answer to the second question can be obtained from the discussion of the previous section: in the used approximation $\cos{x}$ gives the same expression as $\cosh{x}$, similarly with $\sin{x}$ and $\sinh{x}$. Furthermore, when individual terms in $v_z$ change sign, it happens for all of them simultaneously, $v^2_z$ is thus unaffected, and the same applies to $v_c$ and $v_s$. This allows us to conclude that
 \begin{itemize}
		\item for all $t\ne {0,1}$ the negative spectrum is connected to zero only if the same is true for the~positive one.
	\end{itemize}
Regarding the number of negative spectral bands of $H_{\gamma,t}$, our elementary cell has two vertices of degree three with eigenvalues \eqref{Interpol_condition_eigenvalues}. Inspecting their imaginary parts with Theorem~2.6 of \cite{BET22} in mind, we see that
 \begin{itemize}
		\item for negative $\gamma$ there might be at most four distinct negative spectral bands, while a non-negative $\gamma$ allows at most for two of them,
		\item in the limit $t \to 0+$, the maximum number of bands must necessarily decrease (to two for a negative $\gamma$ or to zero if $\gamma \ge 0$).
	\end{itemize}
Since each Floquet component $H_{\gamma,t}(\theta)$ of our operator represents an analytic family with respect to the interpolation parameter $t$ in the sense of \cite{Ka76}, a simple perturbative argument shows that the `superfluous' bands may escape to $-\infty$ in the limit as Figs.~\ref{Interpol_plot_neg_gam} and~\ref{Interpol_plot_Kirch} illustrate. 	

	\subsection*{Degenerate cases: vertices of degree four}

As in the $\delta$ coupling case we want to know what happens when we let either $l_1$ or $l_3$ go to zero and the vertex degree becomes four. The absence of flat bands as well as the behaviour around $k^2 = 0$ are not influenced by such limits. On the other hand, the high-energy spectral asymptotics changes in view of the fact that $P_6$ vanishes in both cases.

As before, we consider first the chain with $l_1=0$, when \eqref{Interpol_asymptotic_t_ne_0} reads
 \begin{equation}\label{Int_high_E_l_1_0}
		\frac{P_5}{l^5} + \frac{P_4}{kl^5} + \frac{{P_c\:{\cos}\:{\theta}+P_s\:{\sin}\:{\theta}}}{k^3l^5} = \mathcal{O}(k^{-2}),
	\end{equation}
with
	\begin{equation*}
		\begin{aligned}
			P_5&= -24l^5\beta_b\sin{kl_2}\,\sin{kl_3},\\
			P_4&= 2l^4(9\beta_d+\beta_e)\sin{k\pi}\,\cos{k\pi},\\
			P_c&\approx -16k^2l^4\beta_c(\sin{kl_2} + \sin{kl_3}),\\
			P_s&\approx 16k^2l^4\beta_f(\sin{kl_2} + \sin{kl_3}),\\
		\end{aligned}
	\end{equation*}
where $\approx$ means up to an error of a lower order in $k$. Bands can obviously occur only in the vicinity of the points $k=\frac{m\pi}{l_2}$ or $k=\frac{m\pi}{l_3}$. Without loss of generality we choose $\frac{m\pi}{l_3}$ and express the trigonometric functions in a $\delta$ neighbourhoods of those points. In analogy with the argument used in the non-degenerate case, we eventually arrive at
	\begin{equation*}
		{\Delta}E_m = \frac{8}{3ll_3}\,\frac{\sqrt{{\beta_c}^2 + {\beta_f}^2}}{|\beta_b|} + \mathcal{O}(m^{-1})
	\end{equation*}
if $l_2$ and $l_3$ is incommensurate; naturally we have to \emph{exclude the case $\gamma=0$} in which $\beta_b$ vanishes. On the other hand, if $\frac{l_2}{l_3}$ is rational and ${\sin}\,{\frac{m{\pi}l_2}{l_3}}=0$ holds for some $m$, the band width expression changes to
	\begin{equation*}
		{\Delta}E_m = \frac{8(l_2+l_3)}{3ll_2l_3}\frac{\sqrt{{\beta_c}^2 + {\beta_f}^2}}{|\beta_b|}+\mathcal{O}(m^{-1});
	\end{equation*}
hence in both situations we conclude that
	\begin{itemize}
		\item in the limit $l_1 \to 0$, the positive spectrum has only bands of asymptotically constant width (at the energy scale, corresponding to the $\mathcal{O}(k^{-1})$ at the momentum scale),
        \item it follows that $P_{\sigma}(H_{\gamma,t}) = 0$ holds if $l_1=0$.
	\end{itemize}
As in the non-degenerate case, this means that the Hausdorff-distance convergence of the spectrum as $t\to 0+$ can be uniform on bounded intervals only.	Concerning the negative part of the spectrum, with reference to \cite{BET22} we can claim that
 \begin{itemize}
		\item in the limit $l_1 \to 0$, there are at most two negative spectral bands for negative $\gamma$, while for $\gamma\ge 0$ the maximum number decreases to one.
	\end{itemize}
	
\smallskip

Passing to the other degenerate case, $l_3 \to 0$, the asymptotic form of the spectral condition is given by relation \eqref{Int_high_E_l_1_0} again, but now with
 \begin{equation*}
		\begin{aligned}
			P_5&= -48l^5\beta_b\sin{kl_1}\,\sin{k\pi}\,\cos{k\pi},\\
			P_4&=2l^4(9\beta_d+\beta_e)\,\sin{k\pi}\:(\cos{kl_1}\,\cos{k\pi}-\sin{kl_1}\,\sin{k\pi}),\\
			P_c&\approx-32k^2l^4\beta_c\,\sin{k\pi}\,\cos{k\pi},\\
			P_s&\approx 32k^2l^4\beta_f\,\sin{k\pi}\,\cos{k\pi}.\\
		\end{aligned}
	\end{equation*}
The rough location of the spectral bands is again determined by vanishing of the highest-order term in $k$, now with the coefficient $P_5$, which leads to three options, namely $k=\frac{m\pi}{l_1}$, $\,k=m$, and $=\frac{(2m-1)}{2}$ with $m\in\mathbb{N}$. Finding the asymptotic expressions of the bands in the same way as above, for bands around $\frac{m\pi}{l_1}$ we have
	\begin{equation*}
		{\Delta}E_m = \frac{8}{3}\frac{1}{l\,l_1}\frac{\sqrt{{\beta_c}^2 + {\beta_f}^2}}{|\beta_b|}+\mathcal{O}(m^{-1}),
	\end{equation*}
again \emph{provided that $\gamma\ne 0$} and irrespective of whether $l_1$ is incommensurable with $l_2$ or not; in the latter case the formula is obtained from a higher-order expansion in $\delta$, but with the same result. If $k=m$, the spectral band width is
 \begin{equation*}
		{\Delta}E_m = \frac{8}{3}\frac{1}{l\,l_1}\frac{\sqrt{{\beta_c}^2 + {\beta_f}^2}}{|\beta_b\,{\cos}\,{ml_1}|}+\mathcal{O}(m^{-1}).
	\end{equation*}
	if $\cos{ml_1}$ is never equal to 0 (meaning that $l_1$ is not a rational multiple of $\pi/2$), while on the other hand
 \begin{equation*}
		{\Delta}E_m = \frac{4}{3\,l}\frac{1}{(\frac{l^2_1}{2}+\frac{{\pi}^2}{2})^{\frac{1}{2}}}\frac{\sqrt{{\beta_c}^2 + {\beta_f}^2}}{|\beta_b|}+\mathcal{O}(m^{-1})
	\end{equation*}
	for some $m$. Finally, if $k=\frac{(2m-1)}{2}$
 \begin{equation*}
		{\Delta}E_m = \frac{2}{9(2m-1){\pi}l^2}\abs{\frac{(9\beta_d+\beta_e)\sqrt{{\beta_c}^2 + {\beta_f}^2}}{{\beta^2_b}\sin{\frac{(2m-1)}{2}l_1}}}+\mathcal{O}(m^{-2}),
	\end{equation*}
	assuming $l_1$ attains value for which $\sin{\frac{(2m-1)}{2}l_1} \ne 0$. Should it not be the case, we~simply go to the higher order of $\delta$ and get
 \begin{equation*}
		{\Delta}E_m = \frac{8}{3}\frac{1}{l\,l_1}\frac{\sqrt{{\beta_c}^2 + {\beta_f}^2}}{|\beta_b|}+\mathcal{O}(m^{-1}).
	\end{equation*}
	\begin{itemize}
		\item In the limit $l_3 \to 0$, the positive spectrum has at most asymptotically constant bands at the energy scale, otherwise they shrink as $\mathcal{O}(m^{-1})$.
		\item Overall for this case we have
            \begin{equation*}
			P_{\sigma}(H_{\gamma,t}) = 0;
		\end{equation*}
		the reasoning behind this argument stays the same, because the majority of possible band widths at the energy scale stays asymptotically constant, and the rest decreases as $\mathcal{O}(m^{-1})$.
	\end{itemize}
	
\subsection*{Interpolation with Kirchhoff coupling}

	We finish this paper by considering the case $\gamma = 0$ which we had to exclude when speaking about degenerate chains. It has common behavior with the $t=1$ case in some sense, as $P_5 = P_3 = P_1 = P_0 = 0$, with all the other polynomials having a much simpler structure. Results obtained earlier for $t \ne 0$, particularly the high energy limit and connection of the spectrum to the energy $k^2=0$, still hold, but the zero value of $\gamma$ gives rise to some peculiarities not seen before.
	First of all,
	\begin{equation*}
		\begin{aligned}
			P_6&= 72l^6(\cos{\frac{{\pi}t}{3}}-1)^2\,{\sin}\,{kl_1}\:{\sin}\,{kl_2}\:{\sin}\,{kl_3},\\
			P_4&= 12l^4(\cos{\frac{{\pi}t}{3}}-1)(\cos{\frac{{\pi}t}{3}}+1)[-4\sin{kl_1}+3\sin{k(l_1+l_2+l_3)}\\
			+&\sin{k(l_1+l_2-l_3)}+\sin{k(l_2+l_3-l_1)}+\sin{k(l_3+l_1-l_2)}],\\
			P_2&= 2l^2(\cos{\frac{{\pi}t}{3}}+1)^2[8\sin{kl_1}-9\sin{k(l_1+l_2+l_3)}\\
			+&\sin{k(l_1+l_2-l_3)}+\sin{k(l_2+l_3-l_1)}+\sin{k(l_3+l_1-l_2)}],\\
			P_c&= -16l^2(\cos{\frac{{\pi}t}{3}}+1)[3k^2l^2(\cos{\frac{{\pi}t}{3}}-1)- (\cos{\frac{{\pi}t}{3}}+1)]\,\,(\sin{kl_2}+\sin{kl_3}),\\
			P_s&= 32kl^3\sqrt{3}\,\sin{\frac{{\pi}t}{3}}\,(\cos{\frac{{\pi}t}{3}}+1)\,(\cos{kl_3}-\cos{kl_2}).
		\end{aligned}
	\end{equation*}
In such a situation, we can find flat bands for $k=m$, $m \in \mathbb{N}$, when both the $P_c$ and $P_s$ vanish. We note that $P_4$ and $P_2$ behave as
 \begin{equation*}
		\begin{aligned}
			P_4 &\propto -4\sin{ml_1}\,\sin^2{ml_3},\\
			P_2 &\propto 12\sin{ml_1}\,\sin{m(2\pi-l_3)}\,\sin{ml_3} + 8 \sin{ml_1}\,\sin^2{ml_3},
		\end{aligned}
	\end{equation*}
and in analogy with the $\delta$ coupling spectral analysis we can conclude that
 \begin{itemize}
		\item if $l_j = \frac{p}{q}\pi$, $j = 1,\, 3$, with comprime $p,\,q \in \mathbb{N}$, then $k^2 = q^2m^2$, $m \in \mathbb{N}$, are energies of flat spectral bands.
	\end{itemize}
For degenerate quantum chains with vertices of degree four, $P_6$~is always zero. The highest order of $k$ present is now $k^4$, and the spectral condition now reads either
	\begin{equation*}
		\sin{k(l_2+l_3)} - (\sin{kl_2}+\sin{kl_3})\,\cos{\theta}+ \mathcal{O}(k^{-1}) = 0
	\end{equation*}
for $l_1 \to 0$, or
 \begin{equation*}
		2\,\sin{k\pi}\,\cos{k(\pi+l_1)} - 2\,\cos{\theta}\,\sin{k\pi}\,\cos{k\pi} + \mathcal{O}(k^{-1})= 0
	\end{equation*}
if $l_3 \to 0$. We have seen exactly the same conditions in the asymptotic regime for $t = 0$; this means that
 \begin{itemize}
		\item for $l_1 = 0$ with $\gamma = 0$ we have $P_{\sigma}(H_{0,t}) = 1$ if $l_3 = \pi$ and $P_{\sigma}(H_{0,t}) = \frac{1}{2}$ otherwise,
        \item for $l_3 = 0$ with $\gamma = 0$ we have $P_{\sigma}(H_{0,t}) = \frac{1}{2}$.
 \end{itemize}

	\appendix
	\section{Appendix}\label{Appendix}
	In order to display all polynomials properly, some of them have been split into parts - these are indicated by an additional subindex over which we sum.
	\begin{center}\small
		$\begin{matrix}
			P_6 & 36l^6(\cos{\gamma(1-t)}+1)(\cos{\frac{{\pi}t}{3}}-1)^2\,{\sin}\,{kl_1}\:{\sin}\,{kl_2}\:{\sin}\,{kl_3}\\
			\midrule\\
			P_5 & -24l^5\,\sin{\gamma(1-t)}\:(\cos{\frac{{\pi}t}{3}}-1)^2\\
			&[{\cos}\,{kl_1}\:{\sin}\,{kl_2}\:{\sin}\,{kl_3}+{\sin}\,{kl_1}\:{\sin}\,{k(l_2+l_3)}]\\
			\midrule\\
			P_{4,1} & -16l^4\sin{kl_1}\\
			&[\cos{\gamma(1-t)}\,(\cos{\frac{{\pi}t}{3}}-1)(\cos{\frac{{\pi}t}{3}}+2)+ 2\,(\cos{\frac{{\pi}t}{3}}-1)(\cos{\frac{{\pi}t}{3}}+\frac{1}{2})] \\
			\midrule\\
			P_{4,2} & 9l^4\sin{k(l_1+l_2+l_3)}\\
			&[\cos{\gamma(1-t)}\,(\cos{\frac{{\pi}t}{3}}-1)(\cos{\frac{{\pi}t}{3}}+3) + 3\,(\cos{\frac{{\pi}t}{3}}-1)(\cos{\frac{{\pi}t}{3}}+\frac{1}{3})]\\
			\midrule\\
			P_{4,3} & l^4[\sin{k(l_1+l_2-l_3)}+\sin{k(l_2+l_3-l_1)}+\sin{k(l_3+l_1-l_2)}] \\
			&[7\,\cos{\gamma(1-t)}\,(\cos{\frac{{\pi}t}{3}}-1)(\cos{\frac{{\pi}t}{3}}+\frac{5}{7}) + 5\,(\cos{\frac{{\pi}t}{3}}-1)(\cos{\frac{{\pi}t}{3}}+\frac{7}{5})]\\
			\midrule\\
			P_{3} & 4l^3\sin{\gamma(1-t)}\,(\cos{\frac{{\pi}t}{3}}-1)(\cos{\frac{{\pi}t}{3}}+1)\\
			&\{8\cos{kl_1}-9\cos{k(l_1+l_2+l_3)}\\
			&-5\,[\cos{k(l_1+l_2-l_3)}+\cos{k(l_2+l_3-l_1)}+\cos{k(l_3+l_1-l_2)}]\}\\
			\midrule\\
			P_{2,1} & -16l^2\sin{kl_1}\\
			&[\cos{\gamma(1-t)}\,(\cos{\frac{{\pi}t}{3}}+1)(\cos{\frac{{\pi}t}{3}}-2)- 2\,(\cos{\frac{{\pi}t}{3}}+1)(\cos{\frac{{\pi}t}{3}}-\frac{1}{2})] \\
			\midrule\\
			P_{2,2} & 9l^2\sin{k(l_1+l_2+l_3)}\\
			&[\cos{\gamma(1-t)}\,(\cos{\frac{{\pi}t}{3}}+1)(\cos{\frac{{\pi}t}{3}}-3) - 3\,(\cos{\frac{{\pi}t}{3}}+1)(\cos{\frac{{\pi}t}{3}}-\frac{1}{3})]\\
			\midrule\\
			P_{2,3} & l^2[\sin{k(l_1+l_2-l_3)}+\sin{k(l_2+l_3-l_1)}+\sin{k(l_3+l_1-l_2)}] \\
			&[7\,\cos{\gamma(1-t)}\,(\cos{\frac{{\pi}t}{3}}+1)(\cos{\frac{{\pi}t}{3}}-\frac{5}{7})  -5\,(\cos{\frac{{\pi}t}{3}}+1)(\cos{\frac{{\pi}t}{3}}-\frac{7}{5})]\\
			\midrule\\
			P_1 & -24l\sin{\gamma(1-t)}\,(\cos{\frac{{\pi}t}{3}}+1)^2\\
			&[{\cos}\,{kl_1}\:{\sin}\,{kl_2}\:{\sin}\,{kl_3}+{\sin}\,{kl_1}\:{\sin}\,{k(l_2+l_3)}]\\
			\midrule\\
			P_0 & 36(\cos{\gamma(1-t)}-1)(\cos{\frac{{\pi}t}{3}}+1)^2\,\sin\,{kl_1}\:\sin\,{kl_2}\:\sin\,{kl_3}\\
			\midrule\\
			P_s & 16l^2\sqrt{3}\,\sin{\frac{{\pi}t}{3}}\\
			&\{[\sin{kl_2}+\sin{kl_3}]\,\sin{\gamma(1-t)}\,[k^2l^2(\cos{\frac{{\pi}t}{3}}+1)+(\cos{\frac{{\pi}t}{3}}-1)]\\
			&+2kl[\cos{kl_3}-\cos{kl_2}][\cos{\frac{{\pi}t}{3}}\,\cos{\gamma(1-t)}+1]\} \\
			\midrule\\
			P_c & -16l^2\{[\sin{kl_2}+\sin{kl_3}]\\
			&[k^2l^2\,\big{(}\cos{\gamma(1-t)}\,(\cos{\frac{{\pi}t}{3}}-1)(\cos{\frac{{\pi}t}{3}}+2) +2\,(\cos{\frac{{\pi}t}{3}}-1)(\cos{\frac{{\pi}t}{3}}+\frac{1}{2})\big{)}\\
			&+\cos{\gamma(1-t)}\,(\cos{\frac{{\pi}t}{3}}+1)(\cos{\frac{{\pi}t}{3}}-2)- 2\,(\cos{\frac{{\pi}t}{3}}+1)(\cos{\frac{{\pi}t}{3}}-\frac{1}{2})]\\
			&-2kl\,[\cos{kl_3}+\cos{kl_2}]\,\sin{\gamma(1-t)}\,(\cos{\frac{{\pi}t}{3}}-1)(\cos{\frac{{\pi}t}{3}}+1)\} \\
		\end{matrix}$
	\end{center}

\subsection*{Data availability statement}

Data are available in the article.

\subsection*{Conflict of interest}

The authors have no conflict of interest.

\subsection*{Acknowledgements}

The work was supported by the Czech Science Foundation within the project 21-07129S.

\end{document}